\documentclass[sigconf]{acmart}
\usepackage{hyperref}

\AtBeginDocument{%
  }

\begin{document}


\title{The Day My Chatbot Changed: Characterizing the Mental Health\\Impacts of Social AI App Updates via Negative User Reviews} 




\author{Sirajam Munira}
\affiliation{%
  \institution{Rensselaer Polytechnic Institute}
  \city{Troy}
  \country{USA}}
\email{munirs@rpi.edu}

\author{Lydia Manikonda}
\authornote{Corresponding author}
\affiliation{%
  \institution{Rensselaer Polytechnic Institute}
  \city{Troy}
  \country{USA}}
\email{manikl@rpi.edu}








\begin{abstract}
  Artificial Intelligence (AI) chatbots are increasingly used for emotional, creative, and social support, leading to sustained and routine user interaction with these systems. As these applications evolve through frequent version updates, changes in functionality or behavior may influence how users evaluate them. However, work on how publicly expressed user feedback varies across app versions in real-world deployment contexts is limited. This study analyzes 210,840 Google Play reviews of the chatbot application \emph{Character AI}, linking each review to the app version active at the time of posting. We specifically examine negative reviews to study how version-level rating trends, and linguistic patterns reflect user experiences. Our results show that user ratings fluctuate across successive versions, with certain releases associated with stronger negative evaluations. Thematic analysis indicates that dissatisfaction is concentrated around recurring issues related to technical malfunctions and errors. A subset of reviews additionally frames these concerns in terms of potential psychological or addiction-related effects. The findings highlight how aggregate user evaluations and expressed concerns vary across software iterations and provide empirical insight into how update cycles relate to user feedback patterns and underscore the importance of stability and transparent communication in evolving AI systems.
\end{abstract}

\ccsdesc[500]{Human-centered computing~Empirical studies in HCI}
\ccsdesc[500]{Computing methodologies~Information extraction}
\ccsdesc[500]{Information systems~Web searching and information discovery}

\keywords{Character AI, User Reviews, Generative AI, AI Chatbots, Human-AI Interaction, AI Update, Negative Reviews}


\maketitle

\section{Introduction}

Generative AI chatbots are increasingly embedded in everyday routines, offering conversation, assistance, and emotional support. As these systems are updated frequently through new app versions, underlying models, features, or interface elements may change across releases. Such changes may influence how users evaluate the application over time. However, it remains unclear how iterative version updates correspond to observable shifts in publicly expressed user evaluations especially with the emerging applications relying on Large Language Model-based architecture. In this study, we focus on differences across released app versions of a social AI chatbot, \emph{Character AI} to study the how these applications are influencing users who are using them for different social needs. We do not track individual users before and after version updates but instead, we examine how publicly expressed evaluations vary across versions available when the reviews were posted. 

Prior work has examined emotional attachment~\cite{rupprechters, xu2025bonding}, dependence~\cite{yankouskaya2025llm,fang2025ai}, and the broader risks and benefits of AI companionship~\cite{yankouskaya2025can}. Some studies document user disappointment following model changes \cite{pataranutaporn2025my}. However, systematic large-scale quantitative analyses examining how version-level changes relate to shifts in user ratings and review language in deployed mobile chatbot applications remain limited, particularly those that combine rating trends with linguistic and thematic analysis of user feedback. To address this gap, we analyze user reviews from the application \emph{Character AI} by linking each review to the active app version at the time of posting. This version-aligned approach allows us to examine how aggregate evaluations vary across successive releases. We analyze version-level rating trends and linguistic patterns using emotional-language markers from LIWC~\cite{tausczik2010psychological}, and thematic patterns from  negative reviews. To maintain analytic focus on dissatisfaction, linguistic and thematic analyses concentrate on reviews with 1- and 2-star ratings (hereafter referred to as negative reviews).

\begin{itemize}
\item [\textbf{RQ1}] How do user ratings vary across successive app versions of \emph{Character AI}?
\item [\textbf{RQ2}] How do linguistic or thematic patterns characterize negative reviews across app versions?
\item [\textbf{RQ3}] What recurring themes emerge in negative reviews regarding perceived changes in functionality, usability, or conversational quality?
\end{itemize}

\section{Dataset and Preprocessing}

We analyzed user reviews of the GenAI-based chatbot application \emph{Character AI}$^{1}$ retrieved from Google Play Store. All publicly available reviews posted from May 2023 up to September 2025 were retrieved and included in our analysis. The dataset consists of 210,840 total reviews, of which 53,927 are negative (1–2 star ratings). Each review includes a rating on a scale of 1 to 5 stars (1 = extremely negative, 5 = extremely positive), review text, timestamp, and the app version active at the time of posting. These fields allowed us to associate each review with the specific released app version available at the time of submission. We focused on Google Play because it provides review text together with version metadata required for version-level alignment. No personally identifiable information was collected or used in this study.

\footnotetext[1]{
https://play.google.com/store/apps/details?id=ai.character.app
}

\section{Methodology}

\subsection{Version Alignment and Update Detection}
To examine how user evaluations varied across releases, versions were ordered sequentially based on their semantic version numbers. This ensured that versions were arranged in the sequence released by the developers. For each version $v_i$, we computed the change in mean rating relative to the immediately preceding version $v_{i-1}$, defined as: \begin{equation} \Delta_i = \bar{R}_i - \bar{R}_{i-1} \end{equation} where, $\bar{R}_i$ denotes the average rating for version $v_i$. We classified a version as a \textit{negative shock version} if $\Delta_i < -0.3$, and as a \textit{positive shock version} if $\Delta_i > 0.3$, referring to both collectively as \textbf{shock versions}. Versions with $|\Delta_i| \leq 0.3$ were labeled as \textbf{non-shock version}. The threshold of 0.3 was selected to capture substantively meaningful shifts in average rating on the 1-to-5 likert scale while avoiding minor fluctuations that commonly occur across releases. For \emph{Character AI}, the average absolute version-to-version rating change was approximately 0.3, making this value an empirically grounded cutoff that distinguishes routine variation from comparatively larger shifts. While prior work does not establish a standardized threshold for version-level rating changes, this data-driven approach provides a consistent criterion for identifying notable deviations. These version-level trends served as the basis for our analysis of rating variation across app versions in RQ1.

\subsection{Sentiment and Behavioral Trend Analysis}
We measured sentiment scores for each review using a lexicon-based sentiment analyzer implemented via spaCy~\cite{honnibal_2020_1212303} with the TextBlob extension. Mean rating trajectories were plotted across version order to examine how aggregate user evaluations varied across successive app versions. Observed declines or fluctuations in these trajectories were interpreted as version-associated shifts in user evaluation patterns, allowing us to assess whether certain releases corresponded to higher levels of negative evaluation. To further examine variation within negative evaluations, we applied the sentiment analyzer to negative reviews and analyzed thematic patterns separately based on sentiment polarity. We then extracted recurring themes from reviews exhibiting positive sentiment and negative sentiment. This stratified approach enabled us to compare how dissatisfaction was expressed across different affective tones within negative ratings.

\subsection{Linguistic Analysis Using LIWC}
To examine linguistic patterns in negative reviews of \emph{Character AI}, we merged each review with its \textit{LIWC output} using the review ID. We focused on selected linguistic categories, including \textit{emo\_anger}, \textit{emo\_sad}, \textit{emo\_neg}, \textit{emo\_pos}, \textit{tone\_neg}, \textit{tone\_pos}, and \textit{swear}. We compared average \textit{LIWC scores} in negative reviews associated with non-shock versus shock versions. This comparison addressed RQ2 by examining whether differences in linguistic markers were associated with version-level rating changes.

\subsection{Thematic Analysis Using $n$-grams}
$n$-grams provide an efficient way to identify frequently occurring word sequences in a text corpus and to surface recurring vocabulary patterns. To examine how negative reviews of \emph{Character AI} referenced app version changes, we extracted $n$-grams from the negative review corpus after removing stopwords. We focused specifically on trigrams ($n$=3) and quadgrams ($n$=4) in this study. These $n$-grams highlighted frequently repeated phrases, such as references to changed behavior (\textit{``bring back old C.AI''}), functional issues (\textit{``please fix app''}), or performance-related concerns. This thematic evidence complemented the LIWC analysis and contributed to RQ3 by identifying recurring patterns in how users described version-associated changes in their reviews.

\subsection{Topic Modeling and Theme Aggregation}
To identify broader thematic structures beyond frequent word patterns, we applied BERTopic~\cite{grootendorst2022bertopic}, a transformer-based topic modeling approach, to the corpus of negative reviews. BERTopic leverages contextual embeddings and density-based clustering to group semantically similar reviews into coherent topics. Each topic was characterized by its most representative terms and sample reviews. To improve interpretability, the resulting topics were further grouped into a smaller set of higher-level themes based on semantic similarity and recurring concepts. Theme aggregation was performed through manual inspection of topic keywords and representative documents to group semantically similar topics. These themes capture dominant categories of user concerns, such as content filtering, login and access issues, monetization, technical performance, and perceived changes in conversational quality. This aggregated thematic structure provided a more interpretable overview of user dissatisfaction and supported RQ3 by identifying recurring themes in how users describe version-associated changes.

\section{Results}

\subsection*{RQ1: User Ratings Across App Versions}

\begin{figure}[h]
  \centering
  \includegraphics[width=\linewidth]{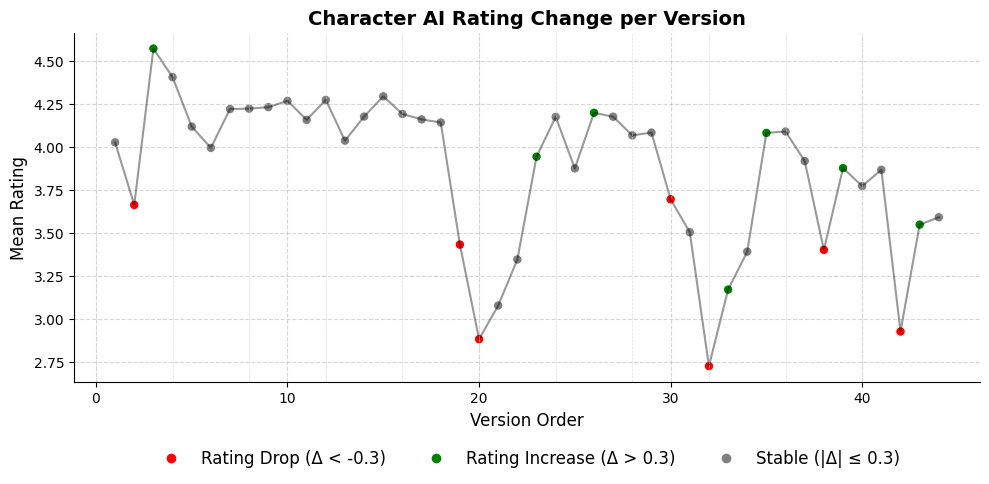}
  \caption{Rating change per version}
  \label{fig:RatingChangeperVersion}
\end{figure}

Version-level trends for \emph{Character AI} (Figure~\ref{fig:RatingChangeperVersion}) show that average user ratings vary across successive app versions. Rather than remaining constant, ratings fluctuate from one version to the next, indicating differences in aggregate user evaluation across releases. These fluctuations include both positive and negative shifts, with several versions exhibiting comparatively larger declines. Notably, negative shifts in average rating are often larger in magnitude than positive shifts. This pattern suggests that certain versions are associated with comparatively stronger negative evaluation. The recurring rise-and-fall cycles indicate that user ratings are sensitive to version-level differences, rather than following a consistent upward or downward trend. Rather than inferring individual-level reactions, these findings demonstrate that publicly expressed evaluations vary across software iterations and highlight the presence of version-associated changes in aggregate user sentiment.

\subsection*{RQ2: Linguistic Patterns in Negative Reviews}

\begin{figure}[h]
  \centering
  \includegraphics[width=\linewidth]{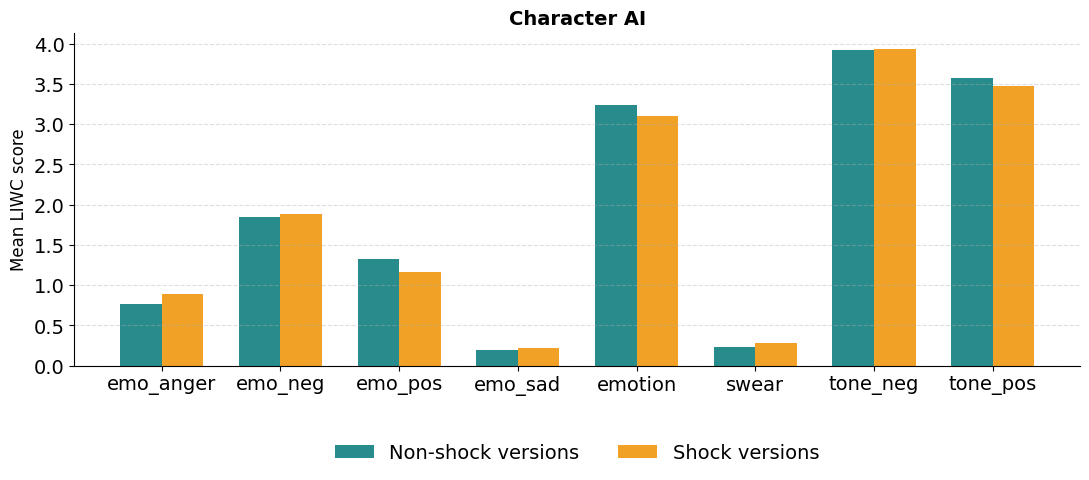}
  \caption{LIWC patterns in negative reviews: Non-shock vs shock versions}
  \label{fig:LIWCpattern}
\end{figure}

Figure~\ref{fig:LIWCpattern} presents the LIWC results comparing linguistic markers in negative reviews across shock and non-shock versions of \emph{Character AI}. Explicit affective markers such as anger or sadness remain relatively low, and differences between shock and non-shock versions are modest in magnitude.
Shock versions are associated with slight increases in negative affect categories (e.g., \textit{anger}, \textit{sadness}) and small decreases in \textit{positive} emotion categories. While these differences indicate variation in linguistic markers across versions, overall affective levels remain comparatively restrained. These findings suggest that version-associated changes are reflected more clearly in rating variation and recurring thematic concerns than in strongly expressed affective language. To further examine how dissatisfaction is articulated, we analyzed thematic patterns within negative reviews stratified by sentiment polarity (Table~\ref{tab:themes}). While both subsets referenced similar core concerns (e.g., \emph{content filtering}, \emph{AI quality decline}, \emph{memory issues}, \emph{monetization pressure}), differences emerged in framing. Reviews with relatively positive sentiment expressed dissatisfaction in a constructive or nostalgic tone (e.g., \textit{``used to love this''}, \textit{``please fix''}), whereas reviews with negative sentiment conveyed stronger hostility, and distrust. 

\begin{table}
  \caption{Dominant themes by sentiment polarity}
  \label{tab:themes}
  \begin{tabular}{ll}
    \toprule
    \textbf{Positive Sentiment} & \textbf{Negative Sentiment}\\
    \midrule
    AI quality decline \& repetition & Aggressive censorship\\
    Memory \& context problems & AI quality regression\\
    Content filter restrictions	& Performance \& server instability \\
    Feature removal/ UX changes & Login \& account access failures \\
    Monetization frustration	&Forced monetization\\
    Performance \& login issues	&Trust \& safety concerns\\
    \bottomrule
  \end{tabular}
\end{table}

\subsection*{RQ3: Thematic Patterns in Negative Reviews}

\begin{figure}[h]
  \centering
  \includegraphics[width=\linewidth]{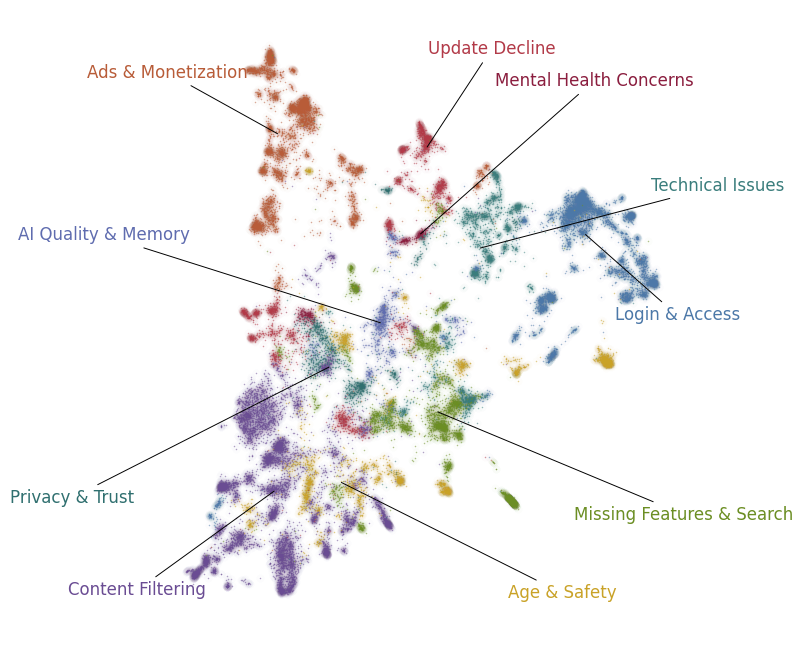}
  \caption{Topic-based visualization of themes expressed in the reviews}
  \label{fig:CharAItheme}
\end{figure}

Figure~\ref{fig:CharAItheme} presents a topic-based visualization of negative review themes derived from BERTopic. The results show that user dissatisfaction is organized around a set of recurring themes, including content filtering and censorship, login and account access issues, monetization (ads and premium features), technical performance problems, missing features, and perceived declines in AI quality and conversational behavior.

These themes are consistent with recurring phrases identified through $n$-gram analysis (Table~\ref{tab:negAttributes}). Expressions such as ``\emph{bring back old C.AI}'', ``\emph{used to love app}'', and ``\emph{please fix app}'' indicate that users frequently compare newer versions with prior experiences. This suggests that dissatisfaction is often framed in relation to perceived changes across versions, rather than isolated issues within a single release. In addition to general update-related dissatisfaction, a subset of negative reviews explicitly references mental-health-related concerns. $n$-gram analysis of reviews containing the term `mental' surfaced recurring phrases such as ``\emph{generally lonely}'', ``\emph{high chance becoming addicted}'', ``\emph{suffer mental illness}'', ``\emph{grown addiction}'', and ``\emph{mental health massive downfall}''. Some reviews also include cautionary language such as ``\emph{please take advice stop using}'' and references to vulnerable groups (e.g., ``\emph{especially teenager}''). These patterns suggest that beyond functional complaints, a portion of users frame their dissatisfaction in terms of perceived psychological risk, addiction, or vulnerability. Overall, the results indicate that negative evaluations are structured around recurring themes related to functionality, usability, and perceived changes in system behavior, with a subset of reviews additionally framing concerns in terms of potential psychological impact.

\begin{table}
  \caption{Top phrases extracted using $n$-grams from all the reviews and a subset of reviews with keyword ``mental''}
  \label{tab:negAttributes}
  \begin{tabular}{cc}
    \toprule
    \textbf{All reviews} & \textbf{includes keyword `mental'}\\
    \midrule
    bring-back-old-c-ai & generally-lonely \\
    used-to-love-app & high-chance-becoming-addicted \\
    hate-new-update & especially-teenager \\
    please-fix-app & suffer-mental-illness \\
    every-time-try & grown-addiction \\
    worse-every-update & mental-health-massive-downfall \\
    get-rid-ads & please-take-advice-stop-using \\
  \bottomrule
\end{tabular}
\end{table}


\subsection{Contributions}
Overall, this study makes four key contributions. First, it introduces a version-level analysis framework that links user reviews to specific app releases and examines how evaluations vary across versions. Second, it provides large-scale empirical evidence that user ratings fluctuate across releases, with certain updates associated with stronger negative evaluations. Third, it combines linguistic, sentiment-based, and topic-based analyses to characterize how dissatisfaction is expressed in negative reviews. Fourth, it identifies recurring themes underlying user dissatisfaction, showing that negative evaluations are structured around functional, usability, and conversational concerns, with a smaller subset referencing perceived psychological or addiction-related risks.

\section{Discussion and Future Work}

This study examined how user evaluations of a GenAI chatbot vary across app versions using large-scale review data. The results show that user ratings fluctuate across successive releases, with certain versions associated with stronger negative evaluation. While explicit affective markers remain relatively stable, thematic analysis indicates that dissatisfaction is consistently expressed through recurring concerns, including content filtering, access issues, monetization, technical performance, and perceived declines in conversational quality. A smaller subset of reviews additionally frames dissatisfaction in terms of potential psychological or addiction-related concerns. These findings suggest that version-level changes are reflected not only in aggregate ratings but also in how users articulate dissatisfaction. Rather than isolated issues, negative evaluations appear to emerge from combinations of functional changes and perceived shifts in system behavior, highlighting the importance of stability and clarity in update practices. 

This study relies on publicly available Google Play reviews, which reflect only users who choose to provide feedback and may not represent the broader population. While version-level alignment links rating variation to specific releases, the effects of version changes cannot be fully isolated from other external factors influencing user evaluations. Our analysis focuses on negative reviews to examine dissatisfaction, including positive and neutral reviews in future work may provide a more complete view of user evaluation dynamics. LIWC and $n$-gram approaches capture explicit patterns but may miss more nuanced expressions. Future work can use embedding-based or qualitative methods, and examine how specific updates relate to changes in user evaluations. Further research could examine how specific updates or release notes correspond to observed shifts in user evaluations, providing a more direct link between system changes and user feedback.

\section{Conclusion}

This study examined how user evaluations of a GenAI chatbot change across app versions using a large-scale dataset of negative user reviews. By linking each review to its corresponding app version, we showed that user ratings do not remain stable but fluctuate across releases, with some versions associated with stronger negative feedback. Our analysis also shows that dissatisfaction is not random. Negative reviews consistently focus on a few recurring issues, including content filtering, login and access problems, ads and monetization, technical performance, and perceived declines in conversational quality and memory. Rather than using strong emotional language, users often express dissatisfaction by comparing newer versions with earlier ones, indicating that expectations are shaped by past experiences. In addition, a smaller group of reviews raises concerns about overuse, dependency, or potential mental health effects, suggesting that user feedback extends beyond functionality to broader user experience concerns. Overall, our results show that version-level changes in AI systems are clearly reflected in both ratings and review content. This highlights that even small updates can influence how users evaluate the system, making consistency and clear communication important in the development of evolving AI applications.


\bibliographystyle{ACM-Reference-Format}
\bibliography{references}

\end{document}